\documentclass{article}
\usepackage[utf8]{inputenc}
\usepackage{multicol}
\usepackage{graphicx}
\usepackage{caption}

\usepackage{xparse}

\title{\huge \textbf{AGD-Autoencoder:} Attention Gated Deep Convolutional Autoencoder for Brain Tumor Segmentation }
\author{Tim Cvetko \\ cvetko.tim@gmail.com}
\date{May 2021}

\begin{document}

\maketitle

\begin{multicols}{2}
\section{Abstract}
\textit{\textbf{
Brain tumor segmentation is a challenging problem in medical image analysis. The endpoint is to generate the salient masks that accurately identify brain tumor regions in an fMRI screening. In this paper, we propose a novel attention gate (AG model) for brain tumor segmentation that utilizes both the edge detecting unit and the attention gated network to highlight and segment the salient regions from fMRI images. This feature enables us to eliminate the necessity of having to explicitly point towards the damaged area(external tissue localization) and classify(classification) as per classical computer vision techniques. AGs can easily be integrated within the deep convolutional neural networks(CNNs). Minimal computional overhead is required while the AGs increase the sensitivity scores significantly. We show that the edge detector along with an attention gated mechanism provide a sufficient enough method for brain segmentation reaching an IOU of 0.78
} The source code is publicly available at: https://github.com/timothy102/Brain-FMRI. }

\hfill \break
\textbf{Keywords}: fMRI, brain tumor segmentation, CNNs, attention gates, autoencoder, segmentation, biomedical image analysis

\section{Introduction}
Early cancer detection is an undeniable global problem. According to the World Health Organization (WHO) [0], it is the second leading cause of death globally. Recent years have seen an increasing use of convolutional neural networks (CNNs) for this task, but most of them use either 2D networks with relatively low memory requirement while ignoring 3D context, or 3D networks exploiting 3D features while with large memory consumption. In addition, existing methods rarely provide uncertainty information associated with the segmentation result. 

Brain segmentation from multi-modal MRi is important for reproducible and accurate measurement of the tumors and thus can assist better diagnosis, prognosis and treatment planning. Deriving from the fact that images have ambiguous boundaries between normal tissues and brain tumors, the robust image segmentation is not an option. Recently, deep learning methods with Convolutional Neural Networks (CNNs) have become the state-of-the-art approaches for brain tumor segmentation (Bakas et al., 2018). Compared with traditional supervised learning methods such as decision trees (Zikic et al., 2012) and support vector machines (Lee et al., 2005), CNNs can learn the most useful features automatically, without the need for manual design and selection of features.
For medical images, uncertainty information of segmentation results is important for clinical decisions as it can help to understand the reliability of the segmentations (Shi et al., 2011) and identify challenging cases necessitating expert review (Jungo et al., 2018). For example, for brain tumor images, the low contrast between surrounding tissues and the segmentation target leads voxels around the boundary to be labeled with less confidence.

Recently, CNNs as a type of discriminative approach have achieved promising results on multi-modal brain tumor segmentation tasks. Havaei et al. (2016) combined local and global 2D features extracted by a CNN for brain tumor segmentation. Although it outperformed the conventional discriminative methods, the 2D CNN only uses 2D features without considering the volumetric context. To incorporate 3D features, applying the 2D networks in axial, sagittal and coronal views and fusing their results has been proposed (McKinley et al., 2016; Li and Shen, 2017; Hu et al., 2018). However, the features employed by such a method are from cross-planes rather than entire 3D space.

DeepMedic (Kamnitsas et al., 2017b) used a 3D CNN to exploit multi-scale volumetric features and further encoded spatial information with a fully connected Conditional Random Field (CRF). It achieved better segmentation performance than using 2D CNNs but has a relatively low inference efficiency due to the multi-scale image patch-based analysis. Isensee et al. (2018) applied 3D U-Net to brain tumor segmentation with a carefully designed training process. Myronenko (2018) used an encoder-decoder architecture for 3D brain tumor segmentation and the network contained an additional branch of variational auto-encoder to reconstruct the input image for regularization. To obtain robust brain tumor segmentation resutls, Kamnitsas et al. (2017a) proposed an ensemble of multiple CNNs including 3D Fully Convolutional Networks (FCN) (Long et al., 2015), DeepMedic (Kamnitsas et al., 2017b), and 3D U-Net (Ronneberger et al., 2015; Abdulkadir et al., 2016). The ensemble model is relatively robust to the choice of hyper-parameters of each individual CNN and reduces the risk of overfitting. However, it is computationally intensive to run a set of models for both training and inference (Malmi et al., 2015; Pereira et al., 2017; Xu et al., 2018).

\subsection{FMRI imaging}
According to []Functional magnetic resonance imaging (fMRI) depicts changes in deoxyhemoglobin concentration consequent to task-induced or spontaneous modulation of neural metabolism. Since its inception in 1990, this method has been widely employed in thousands of studies of cognition for clinical applications such as surgical planning, for monitoring treatment outcomes, and as a biomarker in pharmacologic and training programs. Technical developments have solved most of the challenges of applying fMRI in practice. These challenges include low contrast to noise ratio of BOLD signals, image distortion, and signal dropout. More recently, attention is turning to the use of pattern classification and other statistical methods to draw increasingly complex inferences about cognitive brain states from fMRI data.

\section{Methodology}
\subsection{Image acquisition}
The dataset was acquired thanks to Mateusz Buda and the Brain Tumor Segmentation Dataset 2019 [] []. The data corresponds to 110 patients included in The Cancer Genome Atlas (TCGA) lower-grade glioma collection with at least fluid-attenuated inversion recovery (FLAIR) sequence and genomic cluster data available.

\subsection{Image preprocessing and augmentation}
FMRI images from the dataset were of different sizes and were provided in floating point number format. The preprocessing step involved reshaping the images to 128x128. 
In order to augment the data, we utilized 4 different (PyTorch) transformations: the rotation(shear range = 0.3), multi-modal scaling, cropping and the height-shear range. --- mogoče dodaj še kako, al pa dej before and after

\subsection{Network architecture}
Tumor segmentation was performed using a deep convolutional autoencoder incoorporating two gated networks. 
\subsubsection{Gated Edge Detector Module}

The idea of providing edge information for further processing is not a new idea. Deriving from the one ..., to the Inf-Net module they have shown great results to provide constraints to guide feature extraction for segmentation. 

In order to learn the edge representation, we feed the low-level feature with moderate resolution to the proposed edge attention [EA] module to a convolutional 1x1 filter to produce the filter mapping of the original image. The gated module is trained using the standard Binary Cross Entropy. 

\subsubsection{Embedded Attention Network}

In the context of neural networks, attention is a technique that mimics cognitive attention. The effect enhances the important parts of the input data and fades out the rest -- the thought being that the network should devote more computing power on that small but important part of the data. Which part of the data is more important than others depends on the context and is learned through training data by gradient descent. [Wikipedia == source]

We can explore the benefits of AGs for medical imaging in the context of image segmentation by proposing a grid-based gating that allows attention coefficients to be more specific to local regions. To capture a sufficiently large receptive field and thus semantic contextual information, the feature map-grid is gradually downsampled in standard CNN architectures. In contrast to the localisation model in multi-stage CNNs, AGs progressively suppress feature responses in irrelevant background regions without the requirement to crop a ROI between networks. The gating vector utilizes the contextual information of the AG network to prune lower-level responses in medical imaging. The AG network is incoorporated inside the deep convolutional autoencoder to highlight salient regions that are passed through the skip connection. After the second convolutional block the embedded attention network is placed. In essence, we utilized the AG network to force the intermediate feature maps to be semantically discriminative at each image.

\subsubsection{Deep Convolutional Autoencoder}

The workflow states that the brain fMRI images are first fed into the EA module for edge detection. Afterwards, the images are passed through the encoder.
An autoencoder is a type of artificial neural network used to learn efficient data codings in an unsupervised manner. The aim of an autoencoder is to learn a representation (encoding) for a set of data, typically for dimensionality reduction, by training the network to ignore signal “noise”.
Data-specific: Autoencoders are only able to meaningfully compress data similar to what they have been trained on. Since they learn features specific for the given training data, they are different than a standard data compression algorithm like gzip. So we can’t expect an autoencoder trained on handwritten digits to compress landscape photos.

The encoder is comprised of four convolutional blocks (512, 256, 128 and 64 filters), and the embedded AG network mentioned earlier. These features are fed to the decoder.

The decoder is comprised of four convolutional blocks (64, 128, 256, 512 filters) to produce the output. 

\end{multicols}

\section{Experimental Results}
\subsubsection{Training the network}
We used a k-fold cross-validation method to test the network performance. The approach we used randomly divides the data into 10 approximately equal batches, to provide the randomness factor. This is something also referred to as the record-wise-cross-validation. was implemented to test the generalization capability of the network in medical diagnostics [25].
The generalization capability in clinical practice represents the ability to predict the diagnosis based
on the data obtained from subjects from which there are no observations in the training process.

Therefore, observations from individuals in the training set must not appear in the test set. If this
is not the case, complex predictors can pick up a confounding relationship between identity and
diagnostic status and so produce unrealistically high prediction accuracy [26]. In order to compare the
performance of our network with other state-of-the-art methods, we also tested our network without
k-fold cross-validation (one test). In all the above-mentioned methods, two data portions were used for
the test, two for validation, and six for training. Both datasets, normal and augmented, were tested
using all the methods.
The network was trained using an Adam optimizer, with a mini-batch size equal to 16 and data
shuffling in every iteration. The early-stop condition that affects when the process of network training
will stop corresponds to one epoch. More specifically, it was tuned to finish the training process after
the one epoch, when the loss starts to increase. The regularization factor was set to 0.004, and the
initial learning rate to 0.0004. The weights of the convolutional layers were initialized using a Glorot
initializer, also known as Xavier initializer [27].

\subsubsection{Hyperparameter Tuning}

In Bayesian statistics, a hyperparameter is a parameter of a prior distribution; the term is used to distinguish them from parameters of the model for the underlying system under analysis.

We utilized a manual search to seek the best possible hyperparameters. We included the batch size, the filter size and the learning rate. 

As the entire hyperpameter tuning process would have taken too much power, we trained each possible set of hyperparameters for precisely 1 epoch. 

\begin{center}
 \begin{tabular}{||c c c c c c |||} 
 \hline
  & Learning rate & Filter size & Batch size & K in K-fold & Final BCE \\ [0.2ex] 
   \hline\hline
  Round 1  & 5e-4 & 128 & 32 & 5 & 0.11 \\  
 \hline\hline
   Round 2  & 5e-4 & 256 & 16 & 5 & 0.34 \\  
 \hline\hline
   Round 3  & 1e-3 & 512 & 32 & 10 & 0.11 \\ 
 \hline\hline
   Round 4  & 1e-3 & 256 & 16 & 5 & 0.11 \\  
 \hline\hline
   Round 5  & 1e-2 & 128 & 128 & 10 & 0.22 \\  
 \hline\hline
  Optimal results  & 1e-3 & 512 & 32 & 10 & 0.11 \\  
 \hline\hline
\end{tabular}
\end{center}

\subsection{Results}
Results of the developer AGD Autoencoder are shown in Table 2 and visualized using the confusion matrices, as shown in Figure 3. We trained the model using weighted Focal Loss Binary Cross Entropy. In order to neglect the imbalance of classes and tumors in the database, we have also shown mean, IOU, precision, recall and F1-Score.

Confusion matrices for the subject-wise 10-fold cross-validation approach for testing data from the augmented dataset are shown in Figure 5. The IOU for the testing data stands at 81 percent. 

\begin{center}
 \begin{tabular}{||c c c c c c |||} 
 \hline
  & Accuracy & Precision & Recall & F1-Score & Final BCE \\ [0.2ex] 
 \hline\hline
  Dataset & 81.4 & 77.1 & 86.0 & 82.1 & 0.12 \\  
 \hline
 4- kind augmented dataset & 82.1 & 89.3 & 93.1 & 88.0 & 0.08 \\  
 \hline\hline
\end{tabular}
\end{center}

\section{Conclusion}
In this work we proposed a novel attention gate model that can easily be incoorporated into a deep convolutional decoder-encoder network for brain tumor segmentation. Our deep learning approach eliminates the necessity of object localization and classification.
We applied the attention model to a straightforward fMRI screening and and it showed great results. This was done by using a gated edge detector neural network along with a attention gated deep convolutional autoencoder. Similarly, experimental results demonstrate that the proposed AGs are highly beneficial for tissue/organ identification and localization. 

\hfill \break
Regarding future work, we will consider other dataset augmentation techniques in order to improve the generalizability capability of the network. One of the main improvements will be adjusting the architecture so that it could be used during brain
surgery, classifying and accurately locating the tumor [49]. Detecting the tumors in the operating room
should be performed in real-time and real-world conditions; thus, in that case, the improvement would
also involve adapting the network to a 3D system [50]. By keeping the network architecture simple,
detection in real time could be possible. In future, we will examine the performance of our designed
neural network, as well as improved ones, on other medical images.

\hfill \break
\textbf{Funding}: This research recieved no external funding.\\
\textbf{Conflict of Interest}: The authors declare no conflict of interest.

\section{Acknowledgements}

[0] https://www.ncbi.nlm.nih.gov/pmc/articles/PMC3073717/

[1] Mateusz Buda, AshirbaniSaha, Maciej A. Mazurowski "Association of genomic subtypes of lower-grade gliomas with shape features automatically extracted by a deep learning algorithm." Computers in Biology and Medicine, 2019. image acqusition 

[2] Maciej A. Mazurowski, Kal Clark, Nicholas M. Czarnek, Parisa Shamsesfandabadi, Katherine B. Peters, Ashirbani Saha "Radiogenomics of lower-grade glioma: algorithmically-assessed tumor shape is associated with tumor genomic subtypes and patient outcomes in a multi-institutional study with The Cancer Genome Atlas data." Journal of Neuro-Oncology, 2017.

[3] Tatman, R. (2017, November). R vs. Python: The Kitchen Gadget Test, Version 1. Retrieved December 20, 2017 from https://www.kaggle.com/rtatman/r-vs-python-the-kitchen-gadget-test.

\end{document}